\documentclass[preprint,amssymb,amsmath]{article}

\usepackage{graphicx}
\usepackage{pifont}
\usepackage{textcomp}
\graphicspath{{fig/}}
\usepackage{dcolumn}
\usepackage{bm}
\usepackage{float}
\usepackage{amssymb}
\usepackage{amsmath}
\usepackage{color}
\begin{document}

\title{A short note on  spin pumping theory    \\
with Landau-Lifshitz-Gilbert equation  \\
under quantum fluctuation;   \\
necessity for quantization of localized spin   \\   \    \\}

\author{Kouki Nakata   \\    \    \\
Yukawa Institute for Theoretical Physics, Kyoto University, \\
Kitashirakawa Oiwake-Cho, Kyoto 606-8502, Japan    }


\maketitle

\begin{abstract}
We would like to 
point out the blind spots of the approach
combining the spin pumping theory proposed by Tserkovnyak et al. with the Landau-Lifshitz-Gilbert equation;
this method has been widely used for interpreting  vast experimental results.
 The essence of the  spin pumping effect is  the quantum fluctuation.
Thus, localized spin degrees of freedom should be quantized, 
i.e. be treated as magnons not as classical variables.
Consequently,
the precessing ferromagnet   can be regarded  as a magnon battery.
This point of view will be useful for further progress of spintronics.
\end{abstract}

\section{Introduction}
\label{sec:intro}

A standard way to generate a spin current is
the spin pumping effect at the interface between a ferromagnet and  a non-magnetic material. 
The precessing ferromagnet acts as a source of spin angular momentum
to induce a spin current pumped into a non-magnetic  material, under the ferromagnetic resonance; \textit{spin battery}.\cite{battery}
This methods was first  theoretically proposed by Silsbee et al.\cite{sisbee} and
have been developed  by  Tserkovnyak et al.
Though  Tserkovnyak et al.  have phenomenologically treated the spin-flip scattering processes,\cite{mod2}
now their spin pumping theory has been widely used for interpreting  vast experimental results,\cite{AndoPumping,Kurebayashi,pumping5,pumping6}
in particular by experimentalists.
Thus it will be useful to point out the blind spots of their phenomenological\cite{pumping5} formula 
under  time-dependent transverse magnetic fields (i.e. quantum fluctuations),
for    further progress   of spintronics.
This is the main purpose of this paper.

This manuscript is structured as follows.
We point out  the blind spots of the approach
combining the spin pumping theory proposed by Tserkovnyak et al. with the Landau-Lifshitz-Gilbert (LLG) equation 
in sec. \ref{sec:2}.
We show that, on the basis of the Schwinger-Keldysh formalism,
the essence of the spin pumping effect is quantum fluctuations  in sec. \ref{sec:3}.
We also discuss why the approach combining the  theory by Tserkovnyak et al. with the LLG eq.
cannot appropriately describe the spin pumping effect
and propose an alternative way to capture the dynamics.

In this paper, we use the term, \textit{quantum fluctuations},
to indicate time-dependent transverse magnetic fields.
In addition,
we regard the z-axis as the quantization axis,
and take $\hbar =1$.

\section{Spin Pumping Theory   with  
 Landau-Lifshitz-Gilbert  Equation }
\label{sec:2}

According to the phenomenological\cite{pumping5} spin pumping theory  by Tserkovnyak et al. and their notation,\cite{bauerrev,mod2}
the pumped spin current  ${\mathbf{I}}_{\rm{s{\mathchar`-}pump}}$ reads
\begin{eqnarray}
 {\mathbf{I}}_{\rm{s{\mathchar`-}pump}} = {\rm{G}}_{\perp }^{(\rm{R})} {\mathbf{m}}\times  \dot {\mathbf{m}}
                                                                               +    {\rm{G}}_{\perp }^{(\rm{I})}   \dot {\mathbf{m}},
\label{eqn:e1}                                                                               
 \end{eqnarray}                                                                              
where the dot denotes the time derivative.
 We have taken ${e}=1$, and
 $ {\mathbf{m}}({\mathbf{x}},t)$  denotes a unit vector along the magnetization direction;
 they have treated  $ {\mathbf{m}}({\mathbf{x}},t)$ as  classical variables.
The variable ${\rm{G}}_{\perp }$ is the complex-valued mixing conductance that depends on the material;\cite{yunoki,conductance}
${\rm{G}}_{\perp } =  {\rm{G}}_{\perp }^{(\rm{R})} + i  {\rm{G}}_{\perp }^{(\rm{I})}$.
In addition, they suppose that
the magnetization dynamics of ferromagnets can be described by the  LLG eq.;
\begin{eqnarray}
\dot {\mathbf{m}} = \gamma   {\mathbf{H}}_{\rm{eff}}  \times   {\mathbf{m}}  + \alpha   {\mathbf{m}} \times \dot {\mathbf{m}},
\label{eqn:e2}
\end{eqnarray}
where $ \gamma $ is the gyro-magnetic ratio and
$\alpha $ is the Gilbert damping constant that determines the magnetization dissipation rate.

Here it should be emphasized that
though this Gilbert damping constant, $\alpha $,
was  phenomenologically introduced,\cite{Gilbert}
 it can be derived microscopically
 by considering a whole system including spin relaxation;\cite{tatara2}
 thus the effect of  the exchange coupling to conduction electrons should be considered to
 have  already   been included into this Gilbert damping term.
In addition, 
Rebei et al.\cite{LLG5} have started from a quantum model
and have shown that the Gilbert damping term arises  
only in the limit of small deviations from local equilibrium, where fluctuations are negligible.
In other words,  only in the classical limit,\cite{Lieb}
 the LLG eq., eq. (\ref{eqn:e2}), is an appropriate way to describe the dynamics of the ferromagnet.

The effective magnetic field is set as  
 \begin{eqnarray}
 {\mathbf{H}}_{\rm{eff}}  = (\Gamma (t), 0, B),
  \end{eqnarray}
where $\Gamma (t)$  represents a time-dependent transverse magnetic field.
The  LLG   eq. becomes
\begin{equation}
\begin{pmatrix}
    \dot {\rm{m}}^x \\       \dot {\rm{m}}^y  \\     \dot {\rm{m}}^z   
\end{pmatrix}
= \gamma     
\begin{pmatrix}
  -B{\rm{m}^y}  \\    B{\rm{m}^x} -\Gamma {\rm{m}^z}    \\     \Gamma {\rm{m}^y} 
\end{pmatrix}
+\alpha   
\begin{pmatrix}
   {\rm{m}^y} \dot {{\rm{m}} }^z   -\dot {{\rm{m}} }^y  {\rm{m}^z}  \\ 
      {\rm{m}^z} \dot {{\rm{m}} }^x   -\dot {{\rm{m}} }^z  {\rm{m}^x}  \\     
  {\rm{m}^x} \dot {{\rm{m}} }^y   -\dot {{\rm{m}} }^x  {\rm{m}^y} 
\end{pmatrix}.
\label{eqn:e2-2}
\end{equation}
Eq. (\ref{eqn:e2-2})  is substituted into  ${\rm{I}}_{\rm{s{\mathchar`-}pump}}^z$, eq. (\ref{eqn:e1});
we include the contribution of the Gilbert damping term, which depends on the materials,
up to   ${\cal{O}}(\alpha )$;
$ \alpha  \sim 10^{-3}, 10^{-2} $ for ${\rm{Ni}}_{81}{\rm{Fe}}_{19}$ (metal),\cite{AndoPumping} and 
$ \alpha  \sim 10^{-5} $ for ${\rm{Y}}_{3}{\rm{Fe}}_{5}{\rm{O}}_{12}$ (insulator),\cite{spinwave}
as examples.
Their theory is applicable to both ferromagnetic metals and insulators.\cite{battery}

\label{eqn:e71}

Then the z-component of each term in eq. (\ref{eqn:e1}) becomes
\begin{eqnarray}
({\mathbf{m}}\times  \dot {\mathbf{m}})^z
&=&          \gamma B  [         ({\rm{m}}^x)^2+ ({\rm{m}}^y)^2]   - \gamma \Gamma  {\rm{m}}^x  {\rm{m}}^z           
-   \alpha \gamma \Gamma   [{\rm{m}}^x  {\rm{m}}^y  {\rm{m}}^x + ({\rm{m}}^y)^3  + ({\rm{m}}^z)^2  {\rm{m}}^y]    + {\cal{O}}(\alpha ^2).     \\
 {\dot {{\rm{m}}}}^z
&=&  \gamma \Gamma {\rm{m}}^y  +   
           \alpha  \gamma    \{   B  [({\rm{m}}^x)^2+ ({\rm{m}}^y)^2]    -  \Gamma  {\rm{m}}^x   {\rm{m}}^z   \}         + {\cal{O}}(\alpha ^2).    
\label{eqn:e72}                                            
\end{eqnarray}
Consequently, the z-component of the pumped spin current reads
\begin{eqnarray}
{\rm{I}}_{\rm{s{\mathchar`-}pump}}^z    
&=&          {\rm{G}}_{\perp }^{(\rm{R})}  \Big\{
   \gamma B  [         ({\rm{m}}^x)^2+ ({\rm{m}}^y)^2]   - \gamma \Gamma  {\rm{m}}^x  {\rm{m}}^z           
-   \alpha \gamma \Gamma   [{\rm{m}}^x  {\rm{m}}^y  {\rm{m}}^x + ({\rm{m}}^y)^3  + ({\rm{m}}^z)^2  {\rm{m}}^y] \Big\}       \nonumber    \\
&+&    {\rm{G}}_{\perp }^{(\rm{I})}  \Big\{      \alpha  \gamma    \{   B  [({\rm{m}}^x)^2+ ({\rm{m}}^y)^2]    -  \Gamma  {\rm{m}}^x   {\rm{m}}^z   \}     
+\gamma \Gamma {\rm{m}}^y \Big\}   + {\cal{O}}(\alpha ^2).               \\
  & \stackrel{\Gamma \rightarrow 0}{\longrightarrow }&
 [ {\rm{G}}_{\perp }^{(\rm{R})} +   \alpha  {\rm{G}}_{\perp }^{(\rm{I})}]
   \gamma B  [         ({\rm{m}}^x)^2+ ({\rm{m}}^y)^2].           
\label{eqn:e3}                                   
\end{eqnarray}

At finite temperature,
the magnetization is thermally activated; $\dot {\mathbf{m}}\not=0$.\cite{xiao}
Then the time derivative of the z-component, eq. (\ref{eqn:e72}), means
\begin{eqnarray}
 \dot {\rm{m}}^z 
 & =&  \gamma \Gamma {\rm{m}}^y + \alpha \gamma  \{B [({\rm{m}}^x)^2+ ({\rm{m}}^y)^2]   -\Gamma {\rm{m}}^x  {\rm{m}}^z\}  + {\cal{O}}(\alpha ^2). \label{eqn:e4-4}   \\
 &\stackrel{\Gamma \rightarrow 0}{\longrightarrow } &\alpha \gamma  B [({\rm{m}}^x)^2+ ({\rm{m}}^y)^2].  \label{eqn:e4}   \\
& \not=& 0.                                                                               
\label{eqn:e4-3}                             
\end{eqnarray}
Eqs. (\ref{eqn:e3}), (\ref{eqn:e4}),  and  (\ref{eqn:e4-3}) mean that,
within the framework by  Tserkovnyak et al. with the LLG eq.,
they may  gain spin currents at finite temperature 
if only the magnetic field along the z-axis, $B$, is applied;
\begin{eqnarray}
 {\rm{I}}_{\rm{s{\mathchar`-}pump}}^z  \stackrel{\Gamma \rightarrow 0 (B\not=0)}{\nrightarrow }       0.                                                                                        
\label{eqn:e5}                             
\end{eqnarray}
That is, the approach  combining
the spin pumping theory  by Tserkovnyak et al.  with the LLG eq.  concludes that 
spin currents may be pumped at  finite temperature without time-dependent transverse magnetic fields.

\begin{figure}[h]
\begin{center}
\includegraphics[width=12cm,clip]{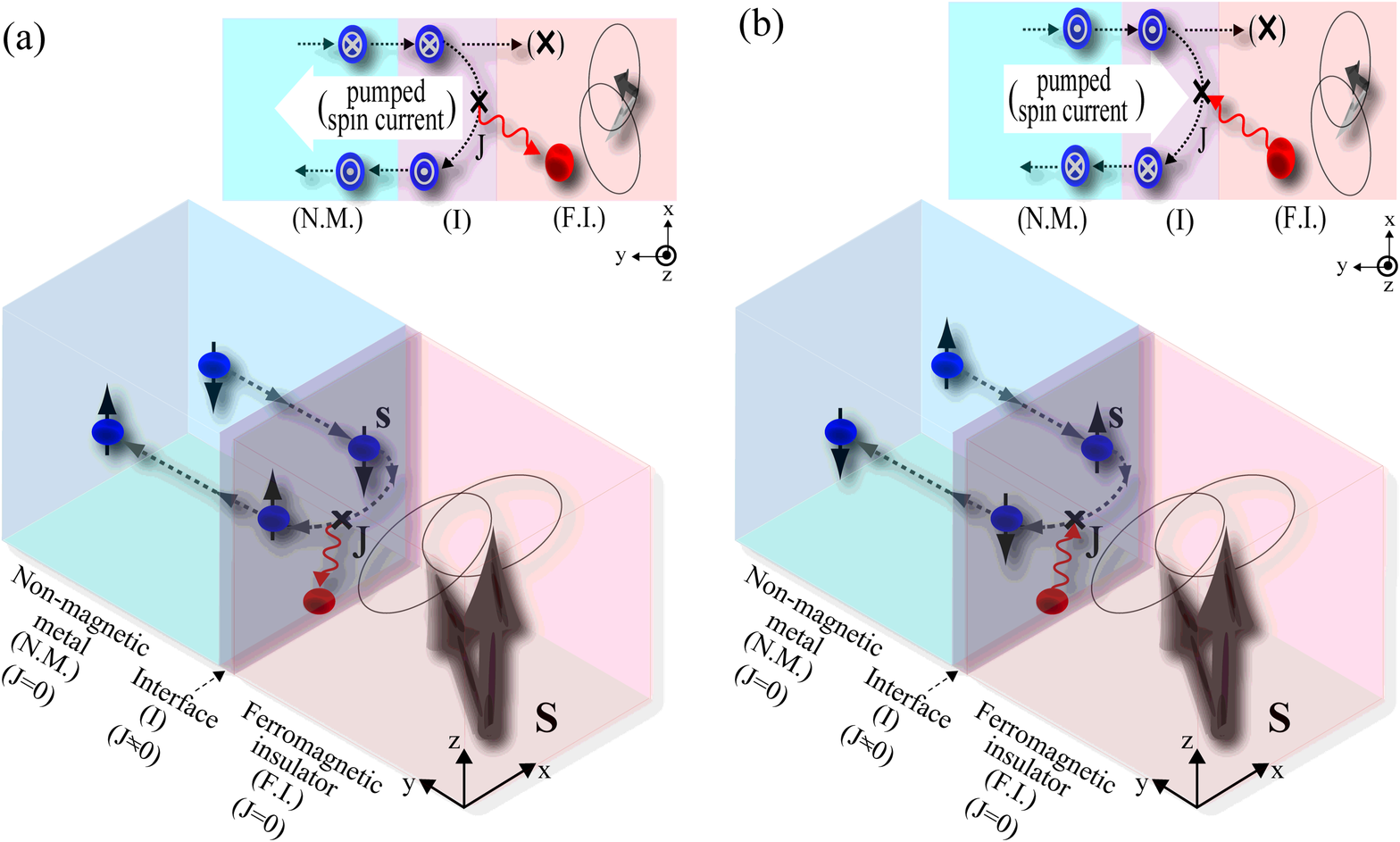}
\caption{(Color online)
A schematic picture of the spin pumping effect mediated by magnons (a), and the inverse process (b).
Spheres  represent magnons and
those with arrows are conduction electrons.
The interface is defined as an effective area where the Fermi gas (conduction electrons) and
the Bose gas (magnons) coexist to interact; $J\not=0$.
Conduction electrons  cannot  enter the ferromagnet,
which is an insulator. 
 \label{fig:pumping} }
\end{center}
\end{figure}

\section{Quantum Spin Pumping Mediated by Magnon   }
\label{sec:3}

\subsection{Spin pumping via the Schwinger-Keldysh formalism }
\label{subsec:kel}

On the other hand,
our spin pumping theory\cite{QSP}  mediated by magnons 
denies the possibility.
We have considered a ferromagnetic insulator and non-magnetic metal junction;
\begin{equation}
 {\cal{H}}_{\rm{ex}}=  - 2Ja_0^3  {\int_{{\mathbf{x}}\in \rm{(interface)}}} d {\mathbf{x}}   {\ }  {\mathbf {S}}({\mathbf{x}},t) \cdot   {\mathbf {s}}({\mathbf{x}},t),
\label{eqn:e9}
\end{equation}
where   $2J$ represents the exchange interaction between localized spins ($ {\mathbf{S}}({\mathbf{x}},t) $, ${\mathbf{x}}=(x, y, z) \in {\mathbf{R}}^3$) and conduction electrons (${\mathbf {s}}$),
and  $a_0$ denotes the lattice constant among ferromagnets (see Fig. \ref{fig:pumping}).
Conduction electron spin variables are represented as  
 \begin{eqnarray}
 {{s}^{j}}  &{=} \sum_{\eta ,\zeta  = \uparrow , \downarrow} [c^{\dagger }_{\eta} (\sigma  ^j)_{\eta \zeta} c_{\zeta}]/2     \\
     &{\equiv}  (c^{\dagger }  \sigma ^j c)/2,
  \end{eqnarray}                  
where $  {  {\sigma }^j} $ are the $ 2\times 2 $  Pauli matrices;
 \begin{eqnarray}
 [ \sigma  ^j , \sigma  ^k  ] = 2i\epsilon _{jkl} \sigma  ^l, (j, k, l = x,y,z).
  \end{eqnarray}
Operators $c^{\dagger }/c $ are   creation/annihilation operators for conduction electrons, 
which satisfy the (fermionic) anticommutation relation;
  \begin{eqnarray}
\{c_{\eta  }({\mathbf{x}}, t), c_{\zeta }^{\dagger }({\mathbf{x}}', t) \}= \delta _{\eta , \zeta } \delta ({\mathbf{x}}-{\mathbf{x'}}).
  \end{eqnarray}
We suppose the uniform magnetization and
thus  localized spin degrees of freedom   can be mapped into magnon ones 
via the Holstein-Primakoff transformation;

We have microscopically  calculated the pumped spin current mediated by magnons
on the basis of the Schwinger-Keldysh formalism\cite{tatara,ramer,kamenev,kita,haug,new} 
with a time dependent transverse magnetic field;
 \begin{eqnarray}
\Gamma (t) = \Gamma _0 {\rm{cos}}(\Omega t), 
 \end{eqnarray}
which acts as a quantum fluctuation.
Our  spin pumping  theory at finite temperature gives\cite{QSP} 
(the concrete calculation has been shown in our previous work\cite{QSP} in detail, and please see it)
\begin{equation}
 {\cal{T}}_{\rm{s}}^z  \stackrel{\Gamma \rightarrow 0 (B\not=0)}{\longrightarrow }   0 + {\cal{O}}(J^2),
\label{eqn:e7}                             
\end{equation}
where ${\cal{T}}_{\rm{s}}$ represents the spin transfer torque;
by integrating over the interface,
the pumped spin current can be estimated.\cite{torqueJMM,QSP}
Thus our formalism   shows that
even when  the magnetic field along the quantization axis (z-axis), $B$, is applied, 
spin currents mediated by magnons cannot flow without quantum fluctuations (i.e. time-dependent transverse magnetic fields).
This is the main difference from  the theory by Tserkovnyak et al. with   the  LLG eq., eq. (\ref{eqn:e5}).
Quantum fluctuations 
induce (net) spin currents, and are essential to spin pumping mediated by magnons 
as well as the exchange interaction;\cite{QSP}
\begin{eqnarray}
 {\cal{T}}_{\rm{s}}^z &  \propto & J{\Gamma_0^2}    \\
 &  \stackrel{\Gamma \rightarrow 0 (B\not=0)}{\longrightarrow }  & 0.
 \label{eqn:e8}                             
\end{eqnarray}
Therefore we call our spin pumping theory
\textit{quantum spin pumping}.\cite{QSP} 

Moreover, it will be useful to mention that
we have revealed that
${\cal{T}}_{\rm{s}}^z$ has a resonance structure as a function
of the angular frequency of the applied transverse field;
the resonance condition reads,
$\Omega = J$,
because the localized spin acts as an effective magnetic field along the quantization axis, $J$,
for conduction electrons. 
This fact (i.e. resonance condition) is useful to enhance the quantum spin pumping effect 
because the angular frequency of a transverse magnetic field is under our control.

In addition, the work by Adachi et al.\cite{adachi,TSP} 
supports  our result;
they have adopted the same approach with ours\cite{QSP} (i.e. the Schwinger-Keldysh formalism with magnon degrees of freedom)
and 
have studied the contribution of the exchange interaction, $J$, to spin pumping up to ${\cal{O}}(J^2)$
without time-dependent transverse magnetic fields (i.e. quantum fluctuations).
They  also suppose the uniform magnetization and
thus  localized spin degrees of freedom   can be mapped into magnon ones 
via the Holstein-Primakoff transformation;
 \begin{eqnarray}
   S^{+} ({\mathbf{x}},t)\equiv S^{x} ({\mathbf{x}},t)+iS^{y} ({\mathbf{x}},t)   \\
=  \sqrt{2\tilde S} a({\mathbf{x}},t) + {\cal{O}}({\tilde S}^{-1/2}),
 \end{eqnarray}                     
 \begin{eqnarray}
   S^{-}({\mathbf{x}},t) \equiv S^{x} ({\mathbf{x}},t)+iS^{y} ({\mathbf{x}},t)  \\
=   \sqrt{2\tilde S}  a^\dagger ({\mathbf{x}},t)    + {\cal{O}}({\tilde S}^{-1/2}),  
 \end{eqnarray}         
    \begin{eqnarray}
S^z({\mathbf{x}},t) = \tilde S-a^\dagger ({\mathbf{x}},t)  a   ({\mathbf{x}},t),
 \end{eqnarray}
      $\tilde S\equiv  S/{a_0^3}$,                             
where operators $a^{\dagger }/a $ are magnon creation/annihilation operators
satisfying the (bosonic) commutation relation; 
 \begin{eqnarray}
 [a(\mathbf{x}, t), a^{\dagger }(\mathbf{x'}, t) ]= \delta (\mathbf{x}-\mathbf{x'}).
  \end{eqnarray}
Up to the $ {\cal{O}}(S)$ terms,
localized spins  reduce to a free boson system.

As the result, the exchange interaction between localized spins and conduction electrons,  
  ${\cal{H}}_{\rm{ex}} (\equiv     {\cal{H}}_{\rm{ex}} ^{S}+{\cal{H}}_{\rm{ex}} ^{\prime})$, 
is rewritten  as
\begin{eqnarray}
 {\cal{H}}_{\rm{ex}} ^{S}   =   -JS  {\int_{{\mathbf{x}}\in \rm{(interface)}}} d    {\mathbf{x}}  
                                                         {\ }   c^{\dagger } ({\mathbf{x}},t)  \sigma^z  c  ({\mathbf{x}},t), 
\label{eqn:e9-2}                                                                                                                                                               
\end{eqnarray}
\begin{eqnarray}
{\cal{H}}_{\rm{ex}} ^{\prime} = -Ja_0^3  \sqrt{\frac{\tilde S}{2}} {\int_{{\mathbf{x}}\in \rm{(interface)}}} d   {\mathbf{x}}  
                                     [a^{\dagger }({\mathbf{x}},t)  c^{\dagger }  ({\mathbf{x}},t) \sigma ^{+} c({\mathbf{x}},t ) 
                                 +  a({\mathbf{x}},t)  c^{\dagger } ({\mathbf{x}},t)  \sigma ^{-} c({\mathbf{x}},t )].
\label{eqn:e12}
\end{eqnarray}
This, ${\cal{H}}_{\rm{ex}} ^{\prime} $, means that 
localized spins lose spin angular momentum by emitting magnons and
conduction electrons flip from down to up by absorbing the momentum 
(Fig. \ref{fig:pumping} (a)), and vice versa (Fig. \ref{fig:pumping} (b)).

 They\cite{adachi,TSP} have  revealed,
through the standard procedure of the Schwinger-Keldysh formalism, 
that  spin currents cannot be generated under the thermal equilibrium condition 
where the temperature difference does not exist between localized spins and conduction electrons,
because of the balance between thermal fluctuations in ferromagnet and those in non-magnetic metal,
 even when a magnetic field along the quantization axis is applied;\cite{adachi,TSP,QSP} 
 \begin{eqnarray}
   {\cal{T}}_{\rm{s}}^z  \stackrel{B \not{=} 0}{= } 0  + {\cal{O}}(J^4).
\label{eqn:e14-2}                         
\end{eqnarray}
 That is,
 a net spin current is induced by 
 inhomogeneous thermal fluctuations between conduction electrons and magnons,
not by the  the applied  magnetic field along the quantization axis  $B$, also in this case.

\subsection{Necessity for the quantization of localized spins; magnon}
\label{subsec:quantization}

 Though the approach  combining the spin pumping theory proposed by Tserkovnyak et al.  with the LLG eq.
has  insisted that spin currents may be pumped at finite temperature 
if only the magnetic field along the z-axis, $B$, is applied (see eq. (\ref{eqn:e5})),
our spin pumping theory mediated by magnons\cite{QSP,adachi,TSP} has denied the possibility 
(see  eqs.  (\ref{eqn:e7}) and (\ref{eqn:e14-2})).
Here  it should be emphasized that 
the Schwinger-Keldysh formalism,\cite{tatara,ramer,kamenev,kita,haug} 
which we have adopted,\cite{QSP,adachi,TSP}
 has microscopically captured the (nonequilibrium) spin-flip dynamics, ${\cal{H}}_{\rm{ex}}^{\prime}$, 
on the basis of the rigorous quantum mechanics beyond phenomenology (see also  Fig. \ref{fig:pumping}).
 In addition as pointed out by Rebei et al.,\cite{LLG5}
though  the LLG eq.  cannot capture the true transient behavior of the system,
the  Schwinger-Keldysh formalism has  microscopically 
described the time development of the spin pumping effect.\cite{QSP,nakatatatara}

 Therefore this result, eqs. (\ref{eqn:e5}), (\ref{eqn:e7}), and (\ref{eqn:e14-2}), 
 means that
the approach combining the spin pumping theory proposed by Tserkovnyak et al. 
with the LLG eq.\cite{LLG5} is unsuited to the description of the dynamics of the spin pumping effect;
due to  the phenomenological treatment of  the spin-flip scattering processes\cite{mod2}
and the classical treatment of localized spin degrees of freedom.

Spin currents are induced by quantum fluctuations.
Then
the LLG eq., which treats localized spins as classical variables not as quantized spinwaves (i.e. magnons),
cannot capture the dynamics appropriately.  
Of course  $  {\mathcal{T}}_{\rm{s}}^z   $ operates the coherent magnon state (as discussed in sec. \ref{subsec:coherent}),
where the uncertainties of the coordinate and of the momentum  is a minimum,\cite{coherent2,swanson}
still, the coherent  magnon state  has been the quantum state.\cite{masahito,LLG5,Lieb}
Thus unless localized spins are quantized, i.e. are treated as magnons,\cite{nakatatatara}
the effect of  quantum fluctuations cannot be included correctly.

 Here it should be noted that as pointed out in sec. \ref{sec:2}, 
 the LLG eq., eq. (\ref{eqn:e2}), is an appropriate way to describe the dynamics of the ferromagnet
 only in the classical limit.\cite{LLG5,Lieb}
Therefore, our approach (i.e. adopting the Schwinger-Keldysh formalism with magnon degrees of freedom)
 is one of the most valid one
to describe the spin pumping effect under time-dependent transverse magnetic fields.

\subsection{Time evolution of the coherent magnon state}
\label{subsec:coherent}

 Last, it will be useful to mention that
the explicit form\cite{adachi,TSP} of  ${\mathcal{T}}_{\rm{s} }^z $  under the Hamiltonian, 
${\cal{H}}_{\rm{ex}} (={\cal{H}}_{\rm{ex}}^{S} +{\cal{H}}_{\rm{ex}} ^{\prime})$, reads
\begin{eqnarray}
 {\mathcal{T}}_{\rm{s}}^z    =        iJa_0^3  \sqrt{\frac{\tilde S}{2}}
                                                          \langle  a^{\dagger }({\mathbf{x}},t)    
                                                                         c^{\dagger }  ({\mathbf{x}},t) \sigma^{+}   c({\mathbf{x}},t )                                                                        
                                                                        -a  ({\mathbf{x}},t)   c^{\dagger } ({\mathbf{x}},t)  \sigma^{-}   c ({\mathbf{x}},t)\rangle,
\label{eqn:e14}                         
\end{eqnarray}
where $\langle \cdot \cdot \cdot \rangle$ denotes the expectation value estimated for the total Hamiltonian.
It is clear that ${\mathcal{T}}_{\rm{s}}^z  $ operates the coherent magnon state.
 According to Glauber et al.,\cite{glauber,mehta2,coherent5}
 if the time derivative of the annihilation operator does not involve a functional dependence on the creation operator, i.e., if
\begin{eqnarray}
       \dot a(t)  = f ( a(t), t),
\label{eqn:e20}                         
\end{eqnarray}
then the states which are initially coherent remain coherent at all times;
that is the coherent state is stable under the time evolution.

Under the Hamiltonian ${\cal{H}}_{\rm{ex}}$ (i.e. eqs. (\ref{eqn:e9-2}) and (\ref{eqn:e12})),
the Heisenberg equation of motion reads 
\begin{eqnarray}
       \dot a({\mathbf{x}},t) =  iJa_0^3  \sqrt{\frac{\tilde S}{2}}     c^{\dagger }  ({\mathbf{x}},t) \sigma^{+}   c({\mathbf{x}},t ).
\label{eqn:e21}                         
\end{eqnarray}
This means that the condition proposed by Glauber et al. is satisfied;
that is the coherent state is stable under the time development.

 In addition,
 also in the case of the quantum spin pumping  under time-dependent transverse magnetic fields,\cite{QSP}
 $  {\mathcal{T}}_{\rm{s}}^z   $  operates the coherent magnon state to have been stable under the time evolution
because it satisfies the generalized  condition proposed by Mista.\cite{mista}

\section{Conclusion}
\label{sec:sum}

 The essence of the  spin pumping effect under  time-dependent transverse magnetic fields
is  not  the applied magnetic field  along the quantization axis (z-axis)
but the quantum fluctuation.
As the result,
localized spin degrees of freedom should be quantized,
i.e. be treated as magnons not as classical variables.
From this  viewpoint,
the precessing ferromagnet  can be regarded  as \textit{magnon battery}.
We consider that the Schwinger-Keldysh formalism
is one of the most valid tools to capture the dynamics appropriately.

\section*{Acknowledgements}

We would like to thank K. Totsuka and S. Onoda
for turning our interest to this issue.
We are supported by the Grant-in-Aid for the Global COE Program "The Next Generation
of Physics, Spun from Universality and Emergence" from the Ministry of Education,
Culture, Sports, Science and Technology (MEXT) of Japan.



\end{document}